\documentclass{emulateapj}
\usepackage{natbib}

\newcount\longrefs
\def\aap{\ifnum\longrefs=1 {Astron.\ Astrophys.}\else 
                           {A\hbox{\rm \&}A}\fi}
\def\aapr{\ifnum\longrefs=1 {Astron.\ Astrophys.\ Rev.}\else 
                            {A\hbox{\rm \&}AR}\fi}
\def\aaps{\ifnum\longrefs=1 {Astron.\ Astrophys.\ Suppl.}\else 
                            {A\hbox{\rm \&}AS}\fi}
\def\aj{\ifnum\longrefs=1 {Astron.\ J.}\else 
                          {AJ}\fi} 
\def\ao{\ifnum\longrefs=1 {Applied Optics}\else 
                           {Appl.\ Opt.}\fi} 
\def\aspcs{\ifnum\longrefs=1 {Astron.\ Soc.\ Pacific Conf. Series}\else 
                           {ASP Conf.\ Ser.}\fi} 
\def\apj{\ifnum\longrefs=1 {Astrophys.\ J.}\else 
                           {ApJ}\fi} 
\def\apjl{\ifnum\longrefs=1 {Astrophys.\ J. Lett.}\else 
                            {ApJ}\fi} 
\def\aplett{\ifnum\longrefs=1 {Astrophys.\ J. Lett.}\else 
                            {ApJ}\fi} 
\def\apjs{\ifnum\longrefs=1 {Astrophys.\ J. Suppl.}\else 
                            {ApJS}\fi}
\def\apss{\ifnum\longrefs=1 {Astrophys.\ and Space Science}\else 
                            {Ap\hbox{\rm \&}SS}\fi}
\def\araa{\ifnum\longrefs=1 {Ann.\ Rev.\ Astron.\ Astrophys.}\else 
                            {ARA\hbox{\rm \&}A}\fi}
\def\azh{\ifnum\longrefs=1 {Astronomicheskii Zhurnal}\else 
                            {Astron.\ Zhur.}\fi}
\def\baas{\ifnum\longrefs=1 {Bull.\ Am.\ Astron.\ Soc.}\else 
                            {BAAS}\fi}
\def\bain{\ifnum\longrefs=1 {Bull.\ Astronom.\ Institutes Netherlands}\else
                            {Bull.\ Astr.\ Inst.\ Neth.}\fi}
\def\gca{\ifnum\longrefs=1 {Geochim.\ Cosmochim.\ Acta}\else 
                           {Geochim.\ Cosmochim.\ Acta}\fi}
\def\grl{\ifnum\longrefs=1 {Geophys.\ Res.\ Lett.}\else 
                           {Geoph.\ Res.\ Lett.}\fi}
\def\iaucirc{\ifnum\longrefs=1 {IAU Circulars}\else 
                          {IAU Circ.}\fi}
\def\ip{\ifnum\longrefs=1 {in press}\else 
                          {in press}\fi}
\def\jchemp{\ifnum\longrefs=1 {J.\ Chem.\ Phys.}\else 
                           {J.\ Chem.\ Phys.}\fi}  
\def\jcp{\ifnum\longrefs=1 {J.\ Chem.\ Phys.}\else 
                           {J.\ Chem.\ Phys.}\fi}  
\def\jgr{\ifnum\longrefs=1 {J.\ Geophys.\ Res.}\else 
                           {J.\ Geophys.\ Res.}\fi}  
\def\jmolspec{\ifnum\longrefs=1 {J.\ Mol.\ Spectrosc.}\else 
                           {J.\ Mol.\ Spectrosc.}\fi}  
\def\jqsrt{\ifnum\longrefs=1 {J.\ Quant.\ Spectrosc.\ Radiat.\ Transfer}\else 
                           {J.\ Quant.\ Spectrosc.\ Radiat.\ Transfer}\fi}  
\def\jrasc{\ifnum\longrefs=1 {J.\ Royal Astron.\ Soc.\ Canada}\else 
                           {JRAS Can.}\fi}  
\def\mnras{\ifnum\longrefs=1 {Mon.\ Not.\ Roy.\ Astron.\ Soc.}\else 
                             {MNRAS}\fi} 
\def\nat{\ifnum\longrefs=1 {Nature}\else 
                           {Nat}\fi}
\def\pasj{\ifnum\longrefs=1 {Pub.\ Astron.\ Soc.\ Japan}\else 
                            {PASJ}\fi} 
\def\pasp{\ifnum\longrefs=1 {Pub.\ Astron.\ Soc.\ Pacific}\else 
                            {PASP}\fi} 
\def\physscr{\ifnum\longrefs=1 {Physica Scripta}\else 
                            {Phys.\ Scrip.}\fi} 
\def\planss{\ifnum\longrefs=1 {Planetary \& Space Science}\else 
                            {Plan. \& Space Sci.}\fi} 
\def\procspie{\ifnum\longrefs=1 {Proc.\ SPIE}\else 
                            {Proc.\ SPIE}\fi} 
\def\qjras{\ifnum\longrefs=1 {Quarterly J.\ Royal Astron.\ Soc.}\else 
                            {QJRAS}\fi} 
\def\sa{\ifnum\longrefs=1 {Soviet Astron..}\else 
                               {Sov.\ Astron.}\fi}
\def\skytel{\ifnum\longrefs=1 {Sky \& Telescope}\else 
                            {Sky \& Tel.}\fi} 
\def\solphys{\ifnum\longrefs=1 {Solar Phys.}\else 
                               {Solar Phys.}\fi}
\def\ssr{\ifnum\longrefs=1 {Space Science Rev.}\else 
                               {Space\ Sci.\ Rev.}\fi}

\def\bibfiles{/home/leen/latex/bibtex/bibliofile}   

\def\apjreferences{\longrefs=0  \bibliographystyle{/home/leen/latex/bibtex/apj}
             \bibliography{/home/leen/latex/bibtex/aajour,\bibfiles}}

\def\dutch{\def\refname{Referenties}\def\abstractname{Samenvatting}%
  \def\bibname{Bibliografie}\def\chaptername{Hoofdstuk}%
  \def\appendixname{Bijlage}\def\contentsname{Inhoudsopgave}%
  \def\listfigurename{Lijst van figuren}\def\listtablename{Lijst van tabellen}%
  \def\indexname{Index}\def\figurename{Figuur}\def\tablename{Tabel}%
  \def\partname{Deel}\def\enclname{Bijlage(n)}\def\ccname{Ter attentie van}%
  \def\headtoname{Aan}\def\headpagename{Pagina}%
  \def\today{\number\day\space\ifcase\month\or januari\or februari\or maart\or%
     april\or mei\or juni\or juli\or augustus\or september\or oktober\or%
     november\or december\fi \space\number\year}%
  \typeout{
              >>>>> use hlatex209 for Dutch hyphenation <<<<< 
         }}
\newcounter{onefig} \newcounter{fignumber}
\newcount\nocaptions \newcount\nofigures \newcount\figwidth
\newcount\viewgraphs
  \def\paper{}  \def\figlabel{} 
\long\def\nextfig#1{\setcounter{figure}{\value{fignumber}}
  \addtocounter{fignumber}{1}
  \ifnum \viewgraphs=1 \newpage \pagestyle{empty} \fi 
  \ifnum\value{onefig}=0 #1 \fi                 
  \ifnum\value{onefig}=\value{fignumber} #1 \fi}
\def\figwidths#1#2{\ifnum \nocaptions=1 #2mm \else #1mm \fi}  
\def\paper#1{}  
\long\def\plotfig#1#2{\ifnum \nofigures=1 \else #2 \fi}
\long\def\captiontext#1{\ifnum \nofigures=1 \raggedright \fi 
   \ifnum \nocaptions=1 \paper
     \ifnum \viewgraphs=0 
       \newline  \mbox{}\hrulefill\mbox{} \newline 
       \newline label:~\{\figlabel\} 
     \fi 
     \else \ifnum \nofigures=0 \fi 
   #1 \fi}

\newcount\panelwidth \newcount\panelheight 
\newcount\bxmin \newcount\bymin \newcount\bxmax \newcount\bymax
\newcount\tbxmin \newcount\tbymin
\newcount\tpanelwidth \newcount\tpanelheight \newcount\tpdif
\def\panelsize #1,#2;{\panelwidth=#1 \panelheight=#2}  
\def\setbb #1,#2;#3,#4;#5,#6;{
  \tbxmin=#1 \tbymin=#2    
  \bxmin=#3 \bymin=#4      
  \bxmax=#5 \bymax=#6}     
\def\barepanel #1{%
  \ifnum\panelheight=0 
    \tpdif=\bymax \advance\tpdif by -\bymin
    \multiply \tpdif by \panelwidth
    \tpanelheight=\tpdif
    \tpdif=\bxmax \advance\tpdif by -\bxmin
    \divide \tpanelheight by \tpdif
  \else \tpanelheight=\panelheight \fi
  \epsfig{file=#1,%
     bbllx=\bxmin bp,bblly=\bymin bp,bburx=\bxmax bp,bbury=\bymax bp,clip=,%
     width=\panelwidth mm,height=\tpanelheight mm}}
\def\labelypanel #1{
  \ifnum\panelheight=0 
    \tpdif=\bymax \advance\tpdif by -\bymin
    \multiply \tpdif by \panelwidth
    \tpanelheight=\tpdif
    \tpdif=\bxmax \advance\tpdif by -\bxmin
    \divide \tpanelheight by \tpdif
  \else \tpanelheight=\panelheight \fi
  \tpdif=\bxmax \advance\tpdif by -\tbxmin
  \tpanelwidth=\panelwidth \multiply \tpanelwidth by \tpdif
  \tpdif=\bxmax \advance\tpdif by -\bxmin
  \divide \tpanelwidth by \tpdif
  \epsfig{file=#1,%
    bbllx=\tbxmin bp,bblly=\bymin bp,bburx=\bxmax bp,bbury=\bymax bp,%
    clip=,width=\tpanelwidth mm,height=\tpanelheight mm}}
\def\labelxpanel #1{%
  \ifnum\panelheight=0 
    \tpdif=\bymax \advance\tpdif by -\bymin
    \multiply \tpdif by \panelwidth
    \tpanelheight=\tpdif
    \tpdif=\bxmax \advance\tpdif by -\bxmin
    \divide \tpanelheight by \tpdif
  \else \tpanelheight=\panelheight \fi
  \tpdif=\bymax \advance\tpdif by -\tbymin
  \multiply \tpanelheight by \tpdif
  \tpdif=\bymax \advance\tpdif by -\bymin
  \divide \tpanelheight by \tpdif
  \epsfig{file=#1,%
    bbllx=\bxmin bp,bblly=\tbymin bp,bburx=\bxmax bp,bbury=\bymax bp,%
    clip=,width=\panelwidth mm,height=\tpanelheight mm}}
\def\labelxypanel #1{%
  \ifnum\panelheight=0 
    \tpdif=\bymax \advance\tpdif by -\bymin
    \multiply \tpdif by \panelwidth
    \tpanelheight=\tpdif
    \tpdif=\bxmax \advance\tpdif by -\bxmin
    \divide \tpanelheight by \tpdif
  \else \tpanelheight=\panelheight \fi
  \tpdif=\bxmax \advance\tpdif by -\tbxmin
  \tpanelwidth=\panelwidth \multiply \tpanelwidth by \tpdif
  \tpdif=\bxmax \advance\tpdif by -\bxmin
  \divide \tpanelwidth by \tpdif 
  \tpdif=\bymax \advance\tpdif by -\tbymin 
  \multiply \tpanelheight by \tpdif
  \tpdif=\bymax \advance\tpdif by -\bymin
  \divide \tpanelheight by \tpdif
  \epsfig{file=#1,%
    bbllx=\tbxmin bp,bblly=\tbymin bp,bburx=\bxmax bp,bbury=\bymax bp,%
    clip=,width=\tpanelwidth mm,height=\tpanelheight mm}}

\def\CC{\par \vspace*{-2ex} \footnotesize \baselineskip=8pt \begin{verbatim}}

\long\def\startignore #1\stopignore{}   

\def\setlistparams{         
  \topsep=0.7ex                 
  \itemsep=0.7ex                
  \leftmargini=3ex}             
\setlistparams                  

\newcounter{alistindex}       

\newcounter{romenumnr}

\newlength{\minipagewidth}

\newsavebox{\boxcontent}
\newcommand{\ovalhead}[1]{
  \unitlength=1cm
  \sbox{\boxcontent}{\mbox{~~{#1}~~}}
  \begin{center}
    \ifdim\wd\boxcontent>6ex 
    \ifdim\wd\boxcontent<8cm 
    \begin{picture}(8,3) \thicklines     
      \put(4.0,0.8){\oval(8,1.6)} 
      \put(0.0,0.7){\parbox{8cm}{
         \begin{center} \usebox{\boxcontent} \end{center}}}
    \end{picture}
    \else \ifdim\wd\boxcontent<12cm 
    \begin{picture}(12,3) \thicklines     
        \put(6.0,0.8){\oval(12,1.6)} 
        \put(0.0,0.7){\parbox{12cm}{
           \begin{center} \usebox{\boxcontent} \end{center}}}
    \end{picture}
    \else
    \begin{picture}(16,3) \thicklines     
        \put(8.0,0.8){\oval(16,1.6)} 
        \put(0.0,0.7){\parbox{16cm}{
           \begin{center} \usebox{\boxcontent} \end{center}}}
    \end{picture}
    \fi \fi \fi
  \end{center}} 

\newcounter{headnr}            
\newcounter{subheadnr}[headnr]
\newcounter{subsubheadnr}[subheadnr]
\def\head #1\par{
  \stepcounter{headnr}                          
  \vspace{2ex} \noindent                        
  {\bf \theheadnr~~~~#1}\\[1ex] \noindent}      
\def\subhead #1\par{  
  \stepcounter{subheadnr}
  \vspace{1.3ex} \noindent
  {\bf \theheadnr.\arabic{subheadnr}~~~#1}\\[0.3ex] \noindent}
\def\subsubhead #1\par{
  \stepcounter{subsubheadnr}
  \vspace{1.0ex} \noindent
  {\bf \theheadnr.\arabic{subheadnr}.\arabic{subsubheadnr}~~~#1}\\ \noindent}

\font\dropfont= cmr12 scaled \magstep5
\def\dropcap#1#2{{\noindent
    \setbox0\hbox{\dropfont #1}\setbox1\hbox{#2}\setbox2\hbox{(}%
    \count0=\ht0\advance\count0 by\dp0\count1\baselineskip
    \advance\count0 by-\ht1\advance\count0by\ht2
    \dimen1=.5ex\advance\count0by\dimen1\divide\count0 by\count1
    \advance\count0 by1\dimen0\wd0
    \advance\dimen0 by.25em\dimen1=\ht0\advance\dimen1 by-\ht1
    \global\hangindent\dimen0\global\hangafter-\count0
    \hskip-\dimen0\setbox0\hbox to\dimen0{\raise-\dimen1\box0\hss}%
    \dp0=0in\ht0=0in\box0}#2}

\def\level #1 #2#3#4{$#1 \: ^{#2} \mbox{#3} ^{#4}$}   

\def\Teff{\hbox{$\rm{T}_{\rm eff}$}}            

\def\cc{\hbox{$\mathrm{^{12}C/^{13}C}$}}    
\def\vt{\hbox{$\xi_t$}}                     

\def\mathstacksym#1#2#3#4#5{\def#1{\mathrel{\hbox to 0pt{\lower 
    #5\hbox{#3}\hss} \raise #4\hbox{#2}}}}

\mathstacksym\lta{$<$}{$\sim$}{1.5pt}{3.5pt} 
\mathstacksym\gta{$>$}{$\sim$}{1.5pt}{3.5pt} 
\mathstacksym\lrarrow{$\leftarrow$}{$\rightarrow$}{2pt}{1pt} 
\mathstacksym\lessgreat{$>$}{$<$}{3pt}{3pt} 

\def\bibfiles{bibliofile} 

\def\apjreferences{\longrefs=0  \bibliographystyle{apj}
             \bibliography{aajour,\bibfiles}}

\slugcomment{Accepted for publication in the {\em{Spitzer}} edition of the
Astrophysical Journal Supplement Series}

\shorttitle{Stellar Models for {\em{Spitzer}}-IRS}
\shortauthors{L. Decin et al.}

\begin{document}


\title{\sc{marcs}-Model Stellar Atmospheres, and Their Application
 to the Photometric Calibration of the {\em{Spitzer}}-IRS
 }


\author{L. Decin\altaffilmark{2}}
\affil{Department of Physics and Astronomy, University of Leuven,
  Celestijnenlaan 200B, B-3001 Leuven (Heverlee), Belgium}
\altaffiltext{1}{Postdoctoral Fellow of the Fund for Scientific Research, 
 Flanders
 {\tt{ email: Leen.Decin@ster.kuleuven.ac.be}}}

\author{P.~W. Morris\altaffilmark{2}, P.~N. Appleton}
\affil{{\em{Spitzer}} Science Center, IPAC, California Institute of Technology,
 M/S 220-6, 1200 E.\ California Blvd., Pasadena CA 91125}
\altaffiltext{3}{NASA {\em{Herschel}} Science Center, IPAC, California 
 Institute of Technology, M/S 220-6, 1200 E.\ California Blvd., 
 Pasadena CA 91125 {\tt{email: pmorris@ipac.caltech.edu}}}
 
\author{V. Charmandaris}
\affil{Cornell University, Astronomy Department, 106 Space Sciences Bldg.,
Ithaca, NY 14853}

\author{L. Armus}
\affil{{\em{Spitzer}} Science Center, IPAC, California Institute of Technology,
 M/S 220-6, 1200 E.\ California Blvd., Pasadena CA 91125}

\and
\author{J.~R. Houck}
\affil{Cornell University, Astronomy Department, 106 Space Sciences Bldg.,
Ithaca, NY 14853}

\setcounter{footnote}{3}

\begin{abstract}
We describe state-of-the-art {\sc{marcs}}-code model atmospheres
generated for a group of A dwarf, G dwarf, and late-G to mid-K giant
standard stars, selected to photometrically calibrate the
{\em{Spitzer}}-IRS\footnote{The IRS was a collaborative venture between 
Cornell University and Ball Aerospace Corporation funded by NASA through the Jet Propulsion 
Laboratory and the Ames Research Center. The {\em{Spitzer}} Space
Telescope is operated by the Jet Propulsion Laboratory,
California Institute of Technology under NASA contract 1407. Support
for this work was provided by NASA through an award issued by
JPL/Caltech.}, and compare the synthetic spectra to observations of 
HR~6688, HR~6705, and HR~7891. The general calibration processes and 
uncertainties are briefly described, and the differences between various 
templated composite spectra of the standards are addressed. In particular, 
a contrast between up-to-date model atmospheres and previously published
composite and synthetic spectra is illustrated for wavelength ranges 
around 8\,$\mu$m (where the SiO $\Delta v =1$ band occurs for the cooler 
standards) and $\lambda \geq 20 \mu$m, where the use of the Engelke function 
will lead to increasingly large discrepancies due to the neglect of gravity 
in cool stars. At this point, radiometric requirements are being met, absolute 
flux calibration uncertainties (1-$\sigma$) are $\sim 20$\,\% in the SH and LH,
and $\sim 15$\,\% in the SL and LL data, and order-to-order
flux uncertainties are $\sim$ 10\,\% or less. 
Iteration between the {\sc marcs} model atmosphere inputs and 
the data processing will improve the S/N ratios and calibration 
accuracies. 
\end{abstract}




\section{Introduction}\label{introduction}

The scientific interpretation and modeling of the spectra produced by the
Infrared Spectrograph \citep[IRS;][]{Houck2004} onboard the {\em{Spitzer}} 
Space Telescope \citep{Werner2004} require 
accurate spectrophotometric calibrations, which depend stars with well known 
environmental and atmospheric properties. 
We have a unique opportunity to use the state-of-the-art {\sc marcs}-code, developed 
by the Uppsala group \citep{Gustafsson1975A&A....42..407G} and modified 
substantially in the 
numerical methods and input line and continuous opacities, to compute 
synthetic spectra of the IRS standard stars.  
The iteration between stellar models and instrument calibrations has been
described by \citet{Decin2000A&A...364..137D, Decin2000d, Decin2000b, Decin2000c}, 
with applications to calibrations of the Short Wavelength Spectrometer (SWS) 
onboard the Infrared Space Observatory (ISO).

In this paper we describe the synthetic spectra based on the
{\sc marcs}-code model atmospheres tailored to a set of standard stars that
are relied upon for the photometric calibration of the
IRS.  We overview the
calibration strategy, and summarize the primary IRS stellar standards.  The 
latest generation model spectra of these
stars are discussed, presented in comparison to IRS spectra of
three standards. We finally contrast our synthetic spectrum for 
$\gamma$~Dra (K5 III) with widely available Kurucz model and composite 
spectra, and summarize the current photometric uncertainties.

\section{The IRS Spectrophotometric Calibration Scheme}\label{calplan}

The basic strategy for spectrally and photometrically calibrating the IRS 
modules has
been outlined by \citet{Morris2003clim.conf..113M}.  Each spectral order
in the Short Low and Long Low (SL and LL, respectively; $\lambda / \Delta\lambda \simeq 
70 - 140$) long slit spectrographs, and the Short High and Long High (SH and LH, 
respectively; $\lambda / \Delta\lambda \simeq$ 700) echelle spectrographs is 
individually calibrated
\subsection{Spectral Calibrations} 
Spectral calibrations are determined by observations of a combination of smooth, 
extended celestial sources, and emission line sources.  
The zodiacal light at its maximum intensity and extended
nebulae pointed off the central source are useful
for determining the order widths, which reflect the widths of the slits, 
spectrally imaged into the arrays.  
The order widths are measured at the 45\%
crossing point of the illumination profile for each order, and are very nearly
5 pixels for each of the SH and LH echelle orders, and 36 pixels for the 
low resolution orders. 
The output signal from the zodiacal light measurements can also be used to 
characterize the pixel-to-pixel response variations along the cross-dispersed 
direction of the SL and LL orders, but is not sufficiently strong for accurately
determining these response variations in the high resolution orders.  
Emission line sources such as Be stars P Cygni (B2pe I) and $\gamma$ Cas (B0.5e IV), 
planetary nebulae NGC7027, NGC6543, SMP083, and SMP031, 
and Saturnian moon Titan all provide strong, unresolved lines to determine
the wavelength solutions, spectral resolutions, and instrument profiles.  
Moving the sources to different locations in the slits allows us
to further characterize impacts on the wavelength calibration by pointing offsets, and 
by pixel undersampling
of the point spread function (PSF).  All four arrays undersample the PSF,
due to the excellent focus of the telescope and of the f/12 light beamed onto the 
IRS arrays, which maximizes the sensitivities.  However, the undersampling
leads to various spectral and photometric complications, such
that (for example) line centroiding accuracies are reliable to $\sim$1/5 a resolution
element, about half of the accuracy achievable for a critically sampled PSF. 
Measurements from flight observations of  
celestial sources supersede the laboratory measurements, 
carried out in pre-launch tests to verify design requirements.

\subsection{Photometric Calibratoins} 
Relative spectral response calibrations are determined by placing photometric 
standards at as many locations in the slits as is feasible (executed as
fine spectral maps), processing and combining
the data collection events into a single 2-D image plane, and removing
signature of the celestial source by means of 
the synthetic spectra.  Ideally the spectral flatfield for each array 
is determined by a weighted mean of flats individually derived from 
different stars; initially they are determined for each array from a 
single star.   The resultant flatfield is then applied to
observations of other standards, and the spectral extractions are analyzed 
to determine the flux conversion coefficients (electrons/second to Janskys)
and polynomial ``tuning'' coefficients to correct for the effects of
diffraction losses on point source observations, and systematic 
low frequency residual errors from the flatfielding.  
The flatfields
are stored as FITS image files to be used in the IRS Basic Calibrated
Data (BCD) Science pipeline, and the flux conversion and tuning coefficients 
are stored in ASCII-format tables for application to 1-D spectral extractions 
in the post-BCD Science pipeline.

The implementation of the above scheme requires staged efforts.  
In the Science Verification (SV) phase only sparse spectral maps of the
calibration stars could be acquired, and depended on zodiacal
light observations to spatially fill in the orders.\footnote{The zodiacal 
light itself is unreliable in the spectral
dimension due to the likely presence of solid state features and
latitudinal variations in dust temperature and grain
properties \citep{Reach2003Icar..164..384R}.} Due to the relatively low
zodiacal fluxes (in SH, LH, and SL-2nd order), the effects of PSF 
undersampling, and space
weather on the detectors, exhaustive observations of the standard stars 
continue to be carried out in order to meet the
radiometric requirement of 5\,\%.  The details of the stellar
atmosphere models are especially important at this stage.

\begin{table}[t!]
\begin{center}
\caption{IRS photometric calibration stars, and derived
  \Teff\ and $\log$ g. Uncertainties are in parentheses 
  \citep[cf. Eq.~(18) in][]{Decin2000A&A...364..137D}. 
\label{calsources}}
\begin{tabular}{llclrr}
\tableline\tableline
Source  & Sp. & Cal.\tablenotemark{a} & Range\tablenotemark{b} &
 \multicolumn{1}{c}{\Teff} & \multicolumn{1}{c}{$\log$ g} \\ 
 & Type &  & & \multicolumn{1}{c}{[K]} & \multicolumn{1}{c}{[cm/s$^2$]} \\
\tableline
HR~6688 & K2~III & P   & SH, LH & 4465 (50)\tablenotemark{c}  & 2.17 (0.19)\tablenotemark{f} \\
HR~7310 & G9~III & P   & [SH],[LH]  & 4830 (50)\tablenotemark{c}  & 2.58 (0.06)\tablenotemark{f} \\
HR~2194 & A0~V   & P   & SL1,SL2    & 10325 (240)\tablenotemark{d} & 4.09 (0.08)  \\
HR~7341 & K1~III & P   & LL2         & 4570 (50)\tablenotemark{c}  & 2.46 (0.08) \\
HR~7950 & A1.5~V & S   & SL1,SL2    & 9060 (120)\tablenotemark{d}  & 3.51 (0.08)\tablenotemark{f}  \\
HR~7891 & A0~V   & S   & SL1,SL2    & 10170 (120)\tablenotemark{d}  & 4.07 (0.07)\tablenotemark{f} \\
HR~6705 & K5~III & S   & SH,LH      & 3980 (50)\tablenotemark{c}    & 1.06 (0.09)\tablenotemark{f} \\
HR~6606 & G9~III & S   & all         & 4975 (50)\tablenotemark{c}& 2.90 (0.07)    \\
HR~2491 & A1~V   & T   & SH,LH      & 10240 (120)\tablenotemark{d} & 4.39 (0.07)\tablenotemark{f}  \\
HD~105  & G0~V   & T   & all         & 5930 (70)\tablenotemark{e} & 4.31 (0.11)\tablenotemark{g}   \\
\tableline
\end{tabular}
\tablenotetext{a}{Calibration types: P\,=\,Principal,
 S\,=\,Secondary,  T\,=\,Testcase. Principal standards are observed to derive flatfields
and absolute flux calibration; secondary standards, which may have limited visibilities 
or less certain stellar parameters, are observed to verify or further refine 
the calibrations.}
\tablenotetext{b}{Entries refer to the IRS modules, as follows: SL1 7.5-14.5 $\mu$m;
SL2 5.3-7.5 $\mu$m; LL1 19.5-38.0 $\mu$m; LL2 14.0-21.3 $\mu$m; SH 9.9-19.6 $\mu$m; 
LH 18.7-37.2 $\mu$m.  Brackets denote where calibration stars may be
used only over certain ranges of the module.} 
\tablenotetext{c}{Derived from ($V-K$).}
\tablenotetext{d}{Derived from ($V-I$).}
\tablenotetext{e}{Derived from ($B-V$)-temperature relation of 
\citet{Flower1996ApJ...469..355F}.}
\tablenotetext{f}{Using [Fe/H] from \citet{Cayrel1997A&AS..124..299C}.}
\tablenotetext{g}{Using [Fe/H] from the $uvby$-$\beta$ -- metallicity
 relation of  \citet{Schuster1989A&A...221...65S}. }
\end{center}
\end{table}

\section{Standard Star Selection}

Table \ref{calsources} summarizes the stars which have been selected and
observed during SV and in IRS Science campaigns for the purposes of 
photometrically calibrating IRS spectroscopy.  The stars were chosen 
in the pre-launch preparatory phase, to meet (as closely as possible) 
specific criteria on the availability of observational data to 
make reasonable estimates of the stellar parameters (described in the next 
section), their environments, and absence of chromospheric 
activity, circumstellar dust shell or disk, multiplicity, or peculiar spectral 
activity.   The IRS operating at low resolution is
$\sim$300 times more sensitive than SWS was at 10 $\mu$m, and consequently
the fainter standard stars may meet the aforementioned requirements to a 
lesser extent than stars in the ISO calibration programs.  We  
make use of observations from the ISO ground-based preparatory program
\citep{Hammersley1998A&AS..128..207H,Hammersley2003clim.conf..129H}
and detailed theoretical work where possible. In particular, HR~6688,
HR~6705, HR~7310, and HR~2491 in Table~\ref{calsources} have been previously 
modeled by \citet{Decin2000d, Decin2000c}
using ground-based and 2.4 -- 12\,$\mu$m ISO-SWS spectroscopy. 
Generally, G-K giants are used to calibrate the LH, LL, and 
SH modules, and the A dwarfs to calibrate SL.  
This balance is struck by the
higher potential for debris disks around the A stars, and
non-photospheric molecular 
and dust layers around late-type M giants. The selection of the
G and K standard stars was nevertheless made on a very restricted basis:
several infrared studies based on the CO $\Delta v = 2$ lines (at
2.3\,$\mu$m) and the CO $\Delta v =1 $ lines (at 4.6\,$\mu$m) have
revealed a thermal dichotomic structure in the outer layers of later-type
stars \citep[e.g.][]{Wiedemann1994ApJ...423..806W}. A two-component
structure --- consisting of the traditional chromosphere and a  radiative
equilibrium region mediated by molecules with CO cooling dropping the
temperature --- with physically distinct areas of 
hot and cool material at the same altitude is thought to better
represent the cool stellar atmosphere than do existing homogeneous
one-component models. For
our purposes, we have selected stars which are thought to belong
to the group of ``quiet'' stars for which the radiatively-cooled
regions very largely dominate the stellar surface, so that the single
component RE atmospheric models are most likely to be valid.







\section{Reference SEDs --- {\sc marcs} synthetic spectra}\label{firstguess}

A new set of theoretical reference SEDs has been calculated using the {\sc
  marcs} and {\sc turbospectrum} code \citep[][and further
  updates]{Gustafsson1975A&A....42..407G, Plez1992A&A...256..551P},
using the same physical input parameters of
  atomic and molecular equilibrium constants, solar
  abundances, continuous opacity sources, etc.,  as the
  ones described by \citet{Decinthesis}. For line
  opacities in the IRS spectral range, a database of infrared lines with
  atomic and molecular transitions (CO, SiO, H$_2$O, OH, NH, HF, HCl, CH,
  and NO) has been prepared. References and discussion
  of the inputs can be found in \citet{Decinthesis}.   
The standard assumptions of homogeneous stationary layers, hydrostatic
  equilibrium and LTE were made. Energy conservation was required for
  radiative and convective flux, where the energy transport due to
  convection was treated through a local mixing-length theory. The
  turbulent pressure was neglected.  The Rosseland optical depth scale
  has been chosen to span values from 6$\times$10$^{-8}$ to 300 in order to
  ensure the fulfillment of the diffusion approximation adopted as
  lower boundary condition for all frequencies, and to minimize the
  the number of frequencies for which the surface layers are
  still optically thick.

\begin{figure}[ht]
\begin{center}
\includegraphics[height=350pt,angle=0]{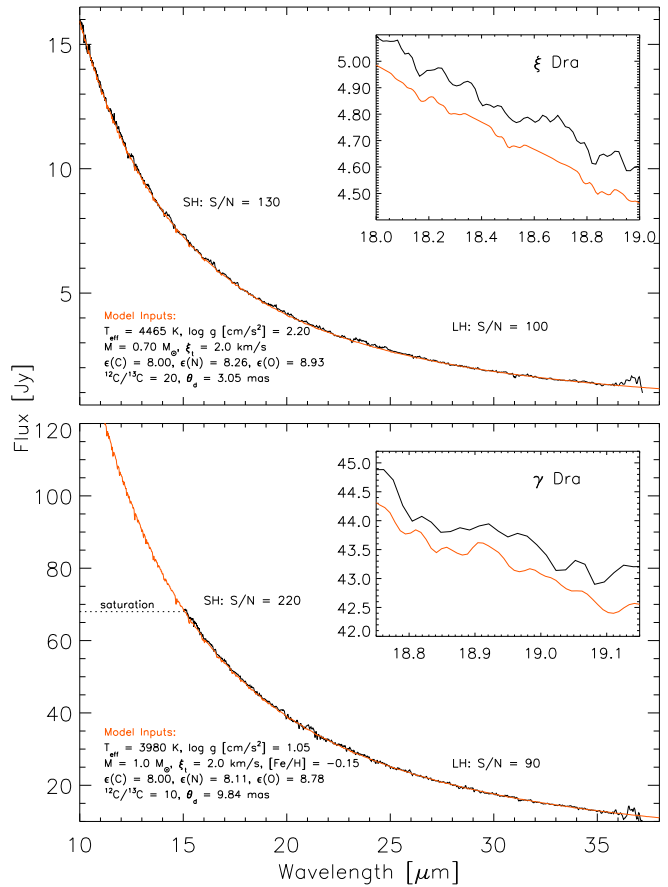}
\caption{High resolution spectra of HR~6688 (top) and 
HR~6705 (bottom), with the {\sc marcs} model SEDs overplotted in red. S/N ratio 
estimates are labelled.  The 
insets show regions of the IRS spectrum around OH lines, compared to the 
model (in red, offset by $-$0.1\,Jy).\label{dra-hi} 
  }
\end{center}
\end{figure}
 
The computed theoretical atmosphere model and
  synthetic spectrum depend on a large number of input parameters, the
  main ones being the effective temperature \Teff, the gravity $g$, the
  microturbulence and the chemical composition. In case of a spherical
  symmetric geometry, one also has to provide either the mass $M$ or the radius
  $R$. the stellar parameters as determined by
  \citet{Decin2000d, Decin2000c} for stars in common with the ISO-SWS 
  calibrators may be adopted.  However, since we aim to set up a
  spectral response calibration independently, we prefer to make initial
  estimates of \Teff and $\log~g$ [cm s$^{-2}$] from
  photometric colors (see below). Since almost no
  information is available for the microturbulence \vt, a value of 2
  km s$^{-1}$ was assumed. Nevertheless, it should be noted that since the final
  calibration of the ISO-SWS relative spectral response functions
  (RSRF) was also based on {\sc marcs} model 
  atmospheres, the IRS spectral response calibration and methods to iterate
  between the input parameters and spectroscopy (e.g., \citet{Decin2000d, Decin2000b, 
  Decin2000c}) are linked to 
  the SWS-RSRFs.  Moreover, first estimates of C, N, and O abundances and the
  \cc-ratio --- being crucial for a correct computation of the molecular 
  opacities in K-giants --- are adopted from recent analyses of the ISO-SWS
  calibrators by \citet{Decin2000d}.
  %
  %
\Teff\ can be calculated for K and M giants directly from $(V-K)$, and for
A-type dwarfs from $(V-I)$, using (semi)-empirical color-temperature
relations \citep[e.g.,][]{Bessell1998A&A...333..231B}. 
Such a calibration is e.g.\ given by
\citet{Bessell1998A&A...333..231B} who derived a polynomial fit
between \Teff\ and $(V-K)$ or $(V-I)$ from (a) the infrared-flux
method (IRFM) for A-K stars, and (b) Michelson interferometry for K-M
giants. For that purpose, the 
$(V-K)$ and $(V-I)$ Johnson or 2MASS
colors were converted to the Johnson-Cousins-Glass system
\citep{Bessell1988PASP..100.1134B} and corrected for interstellar
extinction using $A_V = 0.8$\,mag kpc$^{-1}$
\citep{Blackwell1990A&A...232..396B}, and $E(V-K) = A_V/1.1$ and $A_I
= 0.48 A_V$ \citep{Mathis1990ARA&A..28...37M}, with the distance
calculated from the Hipparcos parallax $\pi$.  Whenever the derived
color estimates were 
outside the ranges specified by \citet{Bessell1998A&A...333..231B} in
their Tables 7 -- 8, the color-temperature relation as determined by
\citet{Flower1996ApJ...469..355F} was used.

In order to estimate the gravity, one needs the radius $R$ and the
stellar mass $M$. The first parameter is assessed from ($K$, BC$_K$,
$\pi$, \Teff($(V-K)_0$)) for K giants or ($V$, BC$_V$, $\pi$,
\Teff($(V-I)_0$)) for A-G dwarfs. The BC$_K$ bolometric corrections
are derived from \citet{Bessell1998A&A...333..231B} whenever
appropriate, otherwise the BC$_V$ data of
\citet{Flower1996ApJ...469..355F} were used. The uncertainty on the
bolometric correction is assumed to be 0.05. M$_{{\rm{bol}},\odot}$ is
assumed to be 4.74 \citep{Bessell1998A&A...333..231B}. Mass
values for the stars in our sample are estimated from evolutionary
tracks with appropriate metallicity as calculated by
\citet{Girardi2000A&AS..141..371G}.  The estimated mass 
depends critically on the assumed metallicity, which has been adopted
from literature references (see Table
\ref{calsources}). Where no information was available, a solar
metallicity was assumed.  The resultant gravity for each star is
listed in Table \ref{calsources}.

\section{IRS Spectra of Representative Standards}\label{irsobs}

In this section we compare reduced spectra of representative standards
to their synthetic SEDs, to assess the models and calibrations.  The
observations have been processed to BCD products using the SSC pipeline 
version S9.5.0, and were response calibrated with the
latest flatfields, except for HR~6688, which was 
calibrated from observations of Sirius and Vega.  
Prior to extraction of low
resolution spectra of HR~7891, the flatfielded BCDs were treated for
removal of background sky emission.  Spectra were extracted and
flux calibrated with the SSC pipeline, from BCDs with the star near
the nod and slit center positions, and then spectra were sigma-clipped
and combined.  
Photometric uncertainties are discussed in Sect.~\ref{prelimerrors}.

\begin{figure}[ht]
\includegraphics[height=140pt,angle=0]{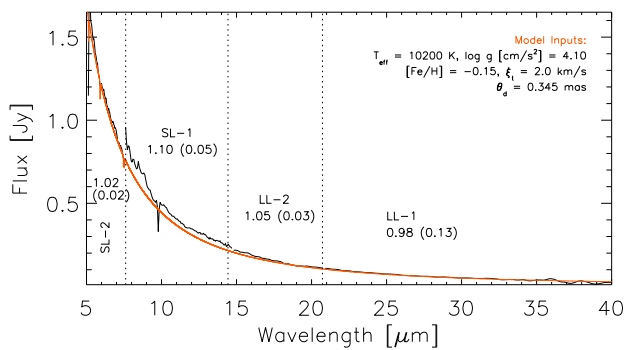}
\caption{Background-corrected low resolution spectrum of HR~7891, 
with the {\sc marcs} model
spectrum overplotted in red. The means and standard deviations
of the ratio between observations and the model SED in each spectral order are
indicated.  The
higher spread of observed LL-1 fluxes about the model is due to the
37.5$-$40\,$\mu$m range where throughput response is very low, and where
a soft filter cutoff allows some contamination with second order light
from $\sim$20\,$\mu$m.  
\label{hr7891lo} 
  }
\end{figure}

\subsection{Bright standards HR~6688 and HR~6705}

The comparison between the IRS spectra and the model of HR~6688
is shown in the upper
panel of Fig.~\ref{dra-hi}.  The data 
quality is sufficiently high to make plausible detections of spectral 
lines with predicted line-to-continuum ratios as low as 1\% (see 
Fig.~\ref{dra-hi} inset displaying OH $\Delta v=0$ spectral features).
The shape of the observed continuum is in excellent agreement with the model
(relying intrinsically on the accuracy of the Sirius model).
The high resolution spectrum of HR~6705, the primary flux calibrator 
for the ISO-SWS, is shown in the lower panel of Fig.~\ref{dra-hi}.  The region 
between 10$-$15\,$\mu$m is not plotted due to saturation over portions of 
echelle orders 15$-$20.  The remaining spectrum is calibrated with the
default calibrations, that is, HR~6688-based flatfields and tuning factors.
The shape of the continuum is again in excellent agreement with the model,  and
the inset plot again demonstrates the potential for detection of weak (molecular) spectral 
features, and improvements to processing methods.




\subsection{HR~7891}

The spectrum of HR~7891 plotted in Fig.~\ref{hr7891lo}, spanning two
orders of magnitude in flux densities, is representative of the
agreement between independently calibrated low resolution spectroscopy
of this A dwarf secondary stardard star, not previously observed at
these wavelengths, and its {\sc marcs} model atmosphere.  The background sky
was subtracted with off-pointed subslits prior to spectral extraction.
The ratio the model
and observations show a $\sim$10\,\% error in the absolute flux calibration of
the SL-1 portion, but otherwise excellent agreement, and that the star
has no detectable thermal excess from a potential debris disk. 

\begin{figure*}[th!]
\begin{center}
\includegraphics[height=500pt,angle=90]{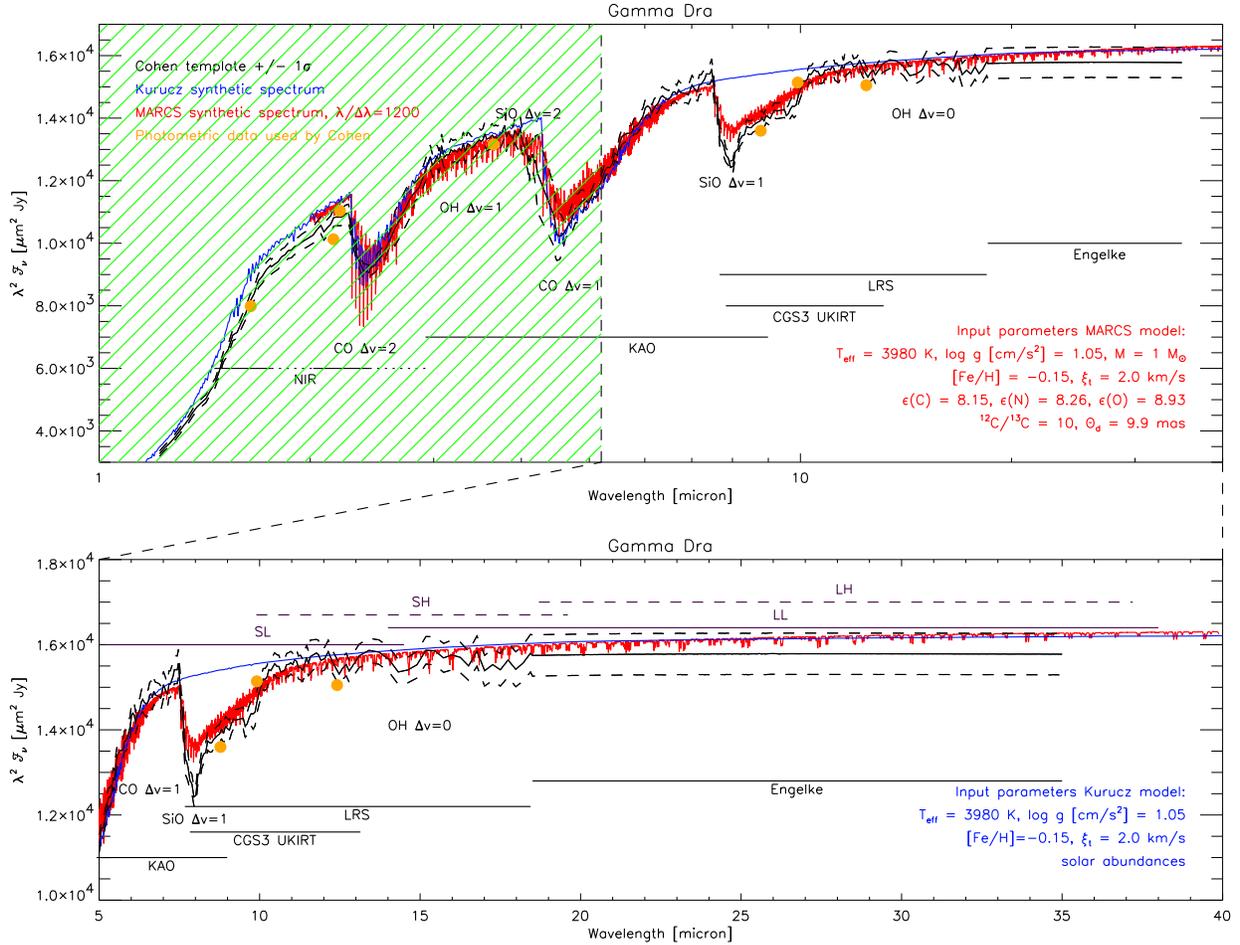}
\caption{Comparison between reference SEDs of HR~6705.
The composite from \citet{Cohen1996AJ....112.2274C} is based on 
different photometric data (orange dots) and spectral fragments spliced 
together. The used data sets are indicated by the solid black line, and
quoted uncertainty limits by the dashed lines. 
The input parameters 
for the MARCS (red) and Kurucz (blue) synthetic spectrum are indicated.
In the bottom panel, the wavelength range observed by the IRS is
enlarged, and the spectral coverage of the SL, LL, SH, and LH are
indicated.
\label{gamdra} 
  }
\end{center}
\end{figure*}

It is helpful to point out the difficulties imposed by background emission,
increasing towards the ecliptic plane to typical levels of several tens
of MJy sr$^{-1}$.  Normally this can be corrected with 
an off-pointing, as done for the observations plotted in Fig.~\ref{hr7891lo}.
If this is not available, then a coarse correction can be applied using a 
{\em{Spitzer}}-centric COBE/DIRBE thermal model of the zodiacal 
dust on the date of the observation, using appropriate scaling factors
for the solid angles of the individual slits.  
The uncertainties associated with the zodiacal light model (including
the likely presence of silicate dust bands), and with the
aperture solid angles, limit the accuracy of the corrections to 
$\sim$20\% at any wavelength. 
A residual tilt may also remain from the fact that both extended and point
source signal may fall through the slits, while calibrations are based solely
on point sources (such that corrections for diffraction losses would not
be adequately compensated by simple scalings using a zodiacal light model).  
 

\section{Comparison with Templated Spectra}\label{compareseds}

The photometric calibration schemes of the ISO and {\em{Spitzer}} instruments have
been partly based on templated datasets, such as those constructed by
M. Cohen and collaborators \citep[e.g.,][]{Cohen2003AJ....125.2645C}, with 
both observational and theoretical components to them. The Cohen datasets can be divided into three
categories: (a) a collection of  
absolutely calibrated photometric and spectroscopic data, merged to create
a continuous spectrum with uncertainties traceable to a specific group of stars
such as $\alpha$ Tau and Vega; (b) a Kurucz model adopted 
at a literature concensus of \Teff, log(g), and metallicity, then fit
to available photometry; and (c) a template constructed from a star
from (a) and/or (b) with the same spectral type, and scaled to the
observed photometric data.  Stars with the same
spectral type may exhibit a different abundance pattern, affecting
e.g.\ the SiO band strengths and continuum levels.  Products in the
(a) group when available are helpful to the IRS since they
rely on no theoretical assumptions, {\em{except}} for the use of an Engelke
function at $\lambda$ $>$ 20 $\mu$m, and some products in the (c) group
are also useful for ``testcase'' standards which can be observed to 
monitor the external photometric stability of the instrument in conjunction 
with pointing performance of the spacecraft, AOT repeatibility, and
statistical measures.  A number of stars not listed in Table~\ref{calsources}
are being observed in normal operations for these activities, and we rely
on products in groups (a) and (c) for stars whose individual value is
their membership in the ensemble.

For the stars listed in Table~\ref{calsources}, we rely explicitly on
the {\sc marcs} synthetic spectra for the detailed calibration
analyses, noting that while a composite is available for HR~6705
\citep{Cohen1996AJ....112.2274C}, it makes use of an Engelke
function extrapolation beyond 18.5 $\mu$m, which we prefer to avoid
(see next paragraph).  For a wider (secondary) set of stars, which
are generally too uncertain in their stellar or environmental parameters
to initially justify the computational resources of MARCS modeling, we can
use composite observations templated by spectral type and luminosity class,
or Kurucz synthetic spectra with the understanding that they are
calculated assuming a plane-parallel geometry, solar abundances
(scaled to the metallicity), and a microturbulent velocity of 2\,km s$^{-1}$,
while these are all free input parameters for the {\sc marcs} atmosphere models.  
Note also that some of the line lists for diatomic molecules used in Kurucz models 
are fifteen years old, incomplete, and do not reproduce line positions very well 
for high $v$ and high $J$.   The stars of this group are typically used for 
trending and checks of order-to-order and module-to-module calibrations, and 
response to high and low flux point sources.


To illustrate these differences over the mid-IR spectral range, we
show three datasets created for HR~6705 in Fig.~\ref{gamdra}.  The
composite constructed by \citet{Cohen1996AJ....112.2274C} is plotted
in black, with the photometric points and the various spectral fragments
indicated.  
Of great importance to the IRS occurs at $\lambda > 18.5$\,$\mu$m, where fluxes are 
approximated with the Engelke function
\citep{Engelke1992AJ....104.1248E}, which offers an analytical
implicit scaling of a semi-empirical solar atmospheric temperature
profile to differing effective temperatures. This analytical
(plane-parallel) approximation should be valid for the 2--60\,$\mu$m continuum 
for giants and dwarfs with effective temperature 3500\,K $\le$ \Teff\ $\le$ 6000\,K, 
where scattering of electrons from H$^-$ dominates the continuum opacity.  
By fitting this function to a set of standard stars,
\citet{Engelke1992AJ....104.1248E} concluded that the estimated
probable error in estimated flux is $\pm 3$\,\% below 10\,$\mu$m, up to
$\pm 5$\,\% in the vicinity of 25\,$\mu$m, and 6\,\% at 60\,$\mu$m. 
The main spectral difference between the {\sc marcs} and Kurucz
theoretical spectra is the absence of the SiO 
fundamental (around 8\,$\mu$m) and first-overtone (around 4\,$\mu$m)
lines in the Kurucz spectrum. Smaller differences do occur due to the use
of a different abundance pattern. With an extension $d =
(R(\tau_{\rm{ross}} = 10^{-5}) /  R(\tau_{\rm{ross}} = 1)) -1 $ being
only 3.7\,\%, the spectral differences between a
plane-parallel and spherical geometry are only marginal. 

The lowermost plot in Fig.~\ref{gamdra} focuses on the wavelength
ranges of the SL, LL, SH, and LH modules. Two main
differences between the 3 data-sets are of major concern for IRS
calibrations: (1) around 8\,$\mu$m where the SiO $\Delta v =1$ lines
occur, and (2) for $\lambda > 18.5$\,$\mu$m.

(1) Comparison to the LRS data indicates that the abundance pattern used as
input for the {\sc marcs} model and synthetic spectrum calculation should be
altered, in particular the oxygen (and carbon) abundance. With the
moderate resolution of IRS ($\sim 600$) and the SL 
mode only starting around 5\,$\mu$m (being in the middle of the CO
$\Delta v =1 $ band for cool giants), it will be very difficult to
constrain the stellar parameters (and in this case more specifically
the oxygen abundance) from the IRS data. This situation can be avoided
by relying on A dwarfs and G giants in
this wavelength range.

(2) At LL and LH wavelengths, we see a clear difference
  between the slope of the {\sc marcs}, Kurucz, and Cohen
  (Engelke) spectrum.  With the Engelke function being a plane-parallel
  approximation, we have compared this function with  {\sc marcs}
  plane-parallel spectra
  for \Teff\ between 3500 and 6000\,K for different values of the
  gravity. The best agreement occurs for \Teff \,=\,6000\,K, which is
  not surprising since Engelke's function is based on a scaled solar
  model. For lower effective temperatures, the role of the gravity
  becomes more important: a
  higher gravity results in a higher opacity (from the free-free
  transitions of H$^-$), a more compact object and
  a smaller temperature gradient due to the requirement of flux
  conservation. As a
  consequence, the infrared continuum slope ($\lambda > 2$\,$\mu$m) is less
  steep for a higher gravity model. From a comparison with the {\sc
  marcs} continua, we may conclude that the uncertainties quoted by
  \citet{Engelke1992AJ....104.1248E} do not account well for the 
  influence of gravity. For HR~6705, the difference in slopes between the 
  {\sc{marcs}} model
  and the composite spectrum (using an Engelke function at $\lambda >$
  18.75~$\mu$m) produces {\em{systematic}} underpredictions of fluxes in
  the composite by 4.2\% at 
  30~$\mu$m, and 6.3\% at 70~$\mu$m.  These differences are easily within 
  grasp of the sensitive IRS and the MIPS detectors.

\section{Summary of the Spectrophotometric Uncertainties}\label{prelimerrors}

First, we estimate that for well-pointed observations of point
sources, the relative photometric uncertainties within any spectral
order generally meet the 5\% radiometric
requirement, over the spectral ranges committed to observers.  The
exception is the two reddest orders of LH (34.2 -- 37.2 $\mu$m), where
throughput response to HR~6688 is low, and will be improved with
HR~6705 calibration observations by the time that this paper is
published.  It must also be noted that the ranges of 
14.1--15.4~$\mu$m in SL 1st order and 38--42.4 $\mu$m in LL 1st
order may be contaminated with light from around 7--8 and 19--21 $\mu$m, 
respectively, due to a weakness in the transmission cutoff of the filter 
in the 2nd spectral orders.  The level of contamination depends on the
color of the source in the slit.

The photometric performance of the IRS is very
sensitive to pointing effects, due to the sizes of the slits and the
PSF widths.  Combined with the effects of PSF undersampling, 
the absolute flux calibration is estimated at this point to be uncertain by 
20\% in the SH and LH data overall, and 15\% in SL and LL data.  Generally, 
order-to-order calibrations are 10\% or less, for point sources in the low
background (or background-corrected) limit\footnote{The low background limit is 
important even for the high resolution 
modules.  Automatic background correction is not performed in the SSC pipeline.} 
and well-placed in the center or nod positions of the slits.  The errors are
easily recognized, and can be mitigated by scaling to photometry, where 
available.

%



\apjreferences




\end{document}